# Mesoporous Silica Nanoparticles Decorated with Polycationic Dendrimers for Infection Treatment

Blanca González,[1,2] Montserrat Colilla,[1,2] Jaime Díez,[1] Daniel Pedraza,[1] Marta Guembe,[1] Isabel Izquierdo-Barba[1,2,]* and María Vallet-Regí[1,2,]*

[1]Departamento de Química Inorgánica y Bioinorgánica, Facultad de Farmacia, Universidad Complutense de Madrid, Instituto de Investigación Sanitaria Hospital 12 de Octubre i+12, Plaza Ramón y Cajal s/n, 28040 Madrid, Spain.

[2]CIBER de Bioingeniería, Biomateriales y Nanomedicina, CIBER-BBN, Madrid, Spain.

* Corresponding authors. E-mail addresses: vallet@ucm.es and ibarba@ucm.es

Phone: +34 91 394 18 61    Fax: +34 91 394 17 86

**Abstract**

This work aims to provide an effective and novel solution for the treatment of infection by using nanovehicles loaded with antibiotics capable of penetrating the bacterial wall, thus increasing the antimicrobial effectiveness. These nanosystems, named "nanoantibiotics", are composed of mesoporous silica nanoparticles (MSNs), which act as nanocarriers of an antimicrobial agent (levofloxacin, LEVO) localized inside the mesopores. To provide the nanosystem of bacterial membrane interaction capability, a polycationic dendrimer, concretely the poly(propyleneimine) dendrimer of third generation (G3), was covalently grafted to the external surface of the LEVO-loaded MSNs. After physicochemical characterization of this nanoantibiotic, the release kinetics of LEVO and the antimicrobial efficacy of each released dosage were evaluated. Besides, internalization studies of the MSNs functionalized with the G3 dendrimer were carried out, showing a high penetrability throughout Gram-negative bacterial membranes. This work evidences that the synergistic combination of polycationic dendrimers as bacterial membrane permeabilization agents with LEVO-loaded MSNs triggers an efficient antimicrobial effect on Gram-negative bacterial biofilm. These positive results open up very promising expectations for their potential application in new infection therapies.

## 1. Introduction

Over the last few years, the development of antimicrobial resistance and its therapeutic implications has become one of the main threats to global health, even talking about the emergence of a new "post-antibiotic era" in which antimicrobial treatment will be useless, given the multidrug-resistant nature of numerous microorganisms [1]. Although the majority of these microorganisms are Gram-positive bacteria, infections caused by the Gram-negative microbes currently have rightly drawn attention, mainly due to the greater difficulty in dealing with them. This drawback lies basically in the presence of an outer membrane that serves as a highly impermeable barrier, as well as additional defense mechanisms that might be absent in Gram-positive bacteria [2-4]. Another added inconvenience is the formation of structures known as *biofilms*, which constitute a natural mechanism of defense of the microorganisms against external aggressions, including antibiotics, reason why their use even hinders the management of these infections [5]. The association of the phenomenon of multiresistance with the appearance of biofilm is results in almost intractable infections using the currently available antibiotic strategies [6].

Nowadays, there is an urgent need to combat biofilm-associated infections through the design of nanoplatforms able to prevent biofilm formation or to produce dispersion of preformed biofilms [7,8]. As well, one of the most important challenges in infection therapy is the design of nanocarriers able to selectively transport antimicrobial agents to the target site and, once there, release them in a controlled fashion [9-12]. Bacteria-targeted nanoparticles offer promising expectations to improve drug efficacy, compliance, and safety [13,14]. Both Gram-positive and Gram-negative bacteria possess different components in their outer membranes, which are easily deprotonated to produce negative charges, increasing their hydrophilicity [15]. Cationic polymers and moieties have been proposed as bacterial detection probes *via* electrostatic interaction [16] and they have been applied in the immobilization and detection of bacteria, mainly owing to the high toxicity of these components [15,17,18].

Dendrimers are single molecular weight, tree-like globular macromolecules with structural uniformity consisting of a central core, branching layers, and numerous end groups [19]. Their unique properties make them good nanoplatforms for different areas of biomedicine, used as protein mimics, drug delivery agents, imaging and antiviral and antibacterial agents [9,20,21]. Moreover, it has been reported that polycationic dendrimers, exhibiting high concentration of positively charged amine groups, are able to efficiently bind to negatively charged bacteria cell walls, increase bacterial membrane permeability and internalize inside the bacteria [10,22].

Regarding inorganic nanomaterials, mesoporous silica nanoparticles (MSNs) are excellent candidates to develop targeted drug delivery devices owing to their high biocompatibility, intrinsically large drug loading capacity and versatility in terms of chemical modification of its surface [23-28]. Herein, a new class of antimicrobial agent, named "nanoantibiotic", based on MSNs functionalized in the external surface with the polypropyleneimine (PPI) dendrimer of third generation (G3) has been developed. Since a generation dependent cytotoxicity has been observed when dealing with polycationic dendrimers, in this approach we have selected a low non-toxic third generation [29]. Moreover, to overcome the drawback of the limited surface charges of low generation dendrimers the attachment of several dendritic molecules to the same nanoparticle is accomplished [30-32]. The positive charge of this nanosystem behaves as internalization agent able to penetrate the negatively charged bacteria wall. The mesoporous structure allows the loading of levofloxacin (LEVO) as antimicrobial drug. The synergistic combination of both elements into a unique nanosystem produces an efficient antimicrobial effect on bacterial biofilm. The influence of the chemical nature of these G3 polycationic dendrimers on the performance of the nanoantibiotic has been investigated. For this purpose, this nanosystem was compared with MSNs externally functionalized with a model low molecular weight diamine. To date accurately mechanistic studies have been reported with star nanoparticles of organic nature (structurally nanoengineered antimicrobial peptide polymers), where membrane attachment and internalization in Gram-negative bacteria have been demonstrated [13]. However, to the best of our knowledge, this is the first study where internalization of inorganic MSNs in Gram-negative bacteria is reported.

## 2. Experimental section

### 2.1. Reagents

All reactions for the synthesis of the silylated dendrimer and for the chemical modification of the silica surface were performed under an inert atmosphere by using standard Schlenk techniques. Dichloromethane ($CH_2Cl_2$, Sigma-Aldrich) was dried by standard procedures over phosphorus (V) oxide and distilled immediately prior to use.

Compound **G3-Si(OEt)$_3$** was synthesized following our previous reported procedure [33]. Fluorescein isothiocyanate (FITC), tetraethylorthosilicate (TEOS), cetyltrimethylammonium bromide (CTAB) and LEVO were purchased from Sigma-Aldrich. 3-Aminopropyltriethoxysilane 97% (APTS), [3-(2-aminoethylamino)propyl]trimethoxysilane 95% (DAMO) and 3-isocyanatopropyltriethoxysilane 95% were purchased from ABCR GmbH & Co. KG., and the G3-PPI dendrimer [G3(NH$_2$)$_{16}$] from SyMO-Chem. These compounds were used without further purification.

Deionized water was further purified by passage through a Milli-Q Advantage A-10 purification system (Millipore Corporation) to a final resistivity of 18.2 MΩ cm. All other chemicals (ammonium nitrate, absolute EtOH, dry toluene, NaOH, etc.) were of the highest quality commercially available and used as received.

The following analytical methods were used to characterize the synthesized compounds: solution state NMR spectroscopy; high resolution magic angle spinning (HR-MAS) NMR spectroscopy; solid state MAS NMR and cross polarization (CP) MAS NMR spectroscopy; Fourier transform infrared spectroscopy (FTIR); elemental chemical analyses; thermogravimetric and differential thermal analysis (TGA); low-angle powder X-ray diffraction (XRD); N$_2$ adsorption porosimetry; electrophoretic mobility measurements to calculate the values of zeta-potential (ζ); dynamic light scattering (DLS); scanning electron microscopy (SEM); energy dispersive X-ray spectroscopy (EDS); and transmission electron microscopy (TEM). The used equipment and conditions are described in the Supporting Information.

## 2.2. Synthetic procedures

### 2.2.1 Synthesis of MSN material

5 µL of APTS (0.023 mmol) was added under stirring and $N_2$ atmosphere to a solution of 2 mg of FITC (0.005 mmol) in 0.250 mL of EtOH. After 2 h stirring at RT in the dark, 5 mL of TEOS (22.4 mmol) was added and the resulting solution was transferred to a syringe dispenser to be used in the next reaction. Separately, 1 g of CTAB (2.74 mmol) was dissolved in 480 mL of water and 3.75 mL of NaOH 2 M and heated to 80 °C. Then, the solution containing the silylated fluorescein derivative and TEOS was added under vigorous stirring at a constant rate of 0.26 mL/min. The vigorous stirring and 80 °C were maintained for 2 h and the suspension was cooled to room temperature (RT). The solid particles were isolated and washed by centrifuging several times with water and EtOH and finally dried. The ion exchange method was employed for the removal of the surfactant using an extracting solution of $NH_4NO_3$ (10 g/L) in $EtOH/H_2O$ (95:5, v/v). A well-dispersed suspension of 1 g of surfactant-containing MSNs in 350 mL of extracting solution was heated to 80 °C under stirring for 2 h and then the solid was thoroughly washed with water and EtOH. This extraction process was repeated two or three times and the solid was dried affording **MSN** material ($S_{BET}$ = 1090 $m^2$/g).

### 2.2.2 Synthesis of the silylated dendrimer G3-Si(OEt)$_3$

Under inert atmosphere, a solution of 3-isocyanotopropyltriethoxysilane (12 µL, 0.018 mmol) in dry $CH_2Cl_2$ (20 mL) was added dropwise to a vigorous stirred solution of G3-PPI dendrimer (0.078 g, 0.018 mmol) in dry $CH_2Cl_2$ (35 mL). The addition was carried out under an inert atmosphere of $N_2$, vigorous stirring and with an addition time of about 2 h. The reaction mixture was stirred overnight at RT and the proper progress of reaction was checked by FTIR and $^1$H NMR spectroscopies. Subsequently, the solution was concentrated by partial removal of solvent under reduced pressure in the vacuum line and the product was immediately used for the covalent attachment to the silica surface.

### 2.2.3. Surface functionalization of the MSN materials

**MSN** materials were firstly dried under vacuum at 80 °C for 5 h and then re-dispersed in a dry solvent under $N_2$, by using vortex shaking and ultrasound cycles for nearly 30 min or until an adequate suspension was attained.

**MSN-DAMO**: A solution of 0.028 g of DAMO (57 μL, 5% excess) in 5 mL of dry toluene was added to a vigorously stirred suspension of 0.500 g of the CTAB-containing MSNs (34% wt, *i.e.*, 0.329 g MSNs) in 100 mL of dry toluene and the mixture was heated to 110 °C overnight. The solid was isolated, exhaustively washed with toluene and acetone and then dried under vacuum at 30 °C. Finally, the surfactant was removed by ion exchange as above described.

**MSN-G3**: For the covalent anchorage of the dendrimers onto the external surface of the MSN sample, the freshly obtained solution of silylated G3-PPI dendrimer, **G3-Si(EtO)$_3$**, was checked (FTIR and $^1$H NMR spectroscopy) and immediately employed in the post-grafting reaction with the MSN material. After solvent evaporation from the solution to approximately 10 mL it was added dropwise under $N_2$ to a suspension of 0.250 g of MSN in 25 mL of dry $CH_2Cl_2$. The mixture was stirred overnight, the solid particles isolated and exhaustively washed with $CH_2Cl_2$ and dried under vacuum at 30 °C. $^1$H HR MAS NMR $\delta_H$ (500 MHz, $D_2O$): 3.27 (br, C*H$_2$*NHCONH), 3.20 (br, C*H$_2$*NH$_2$), 2.86 (br, C*H$_2$*N), 2.07 (br, N*H$_2$*), 1.92 (br, CH$_2$C*H$_2$*CH$_2$), 1.81 (br, CH$_2$C*H$_2$*CH$_2$CH$_2$), 0.87 (t, C*H$_2$*Si).

### 2.3. Drug loading and "in vial" release assays

#### 2.3.1. Levofloxacin loading into MSN materials

LEVO loading in the inner pores of MSN materials was performed following an impregnation method [34]. **MSN** and **MSN-DAMO** materials were loaded by soaking 50 mg of the nanoparticles in 10 mL of a LEVO solution in ethanol (5 mg/mL) and the suspension was stirred at RT for 24 h preserved from light. Then, the materials were filtered, gently washed with absolute ethanol, dried under vacuum and denoted as **MSN-L** and **MSN-DAMO-L** samples. Taking into account the size of the third generation dendrimer, the loading process was performed during the same step of external functionalization of the MSN material with **G3-Si(OEt)$_3$**, which is accomplished in dry $CH_2Cl_2$. In a first assay, MSN material was loaded by soaking 50 mg of the nanoparticles in 10 mL of a LEVO solution in $CH_2Cl_2$ (5 mg/mL) and the suspension was stirred at RT for 16 h in dark conditions. Then, the sample was filtered, gently washed with $CH_2Cl_2$, dried under vacuum and characterized to assess the right loading of LEVO using this solvent. To afford **MSN-G3-L**, 250 mg of vacuum dried MSN were resuspended in 25 mL of a LEVO solution in dry $CH_2Cl_2$ (10 mg/mL) and stirred at RT for 16 h in darkness conditions. To this suspension a solution of the freshly prepared **G3-Si(OEt)$_3$** (amounts and conditions as described above for the functionalization of 250 mg of MSN) was added and stirring was maintained for another 16 h, at RT in the absence of light. Then, the sample was filtered, gently washed with $CH_2Cl_2$ and dried under vacuum. The amount of drug loaded in MSN materials was determined from the elemental chemical analyses and TGA.

### 2.3.2. *"In vial"* levofloxacin release assays from MSNs

The kinetic studies of LEVO release were carried out in phosphate buffered saline (PBS) at 37 °C and physiological pH of 7.4. A double-chamber cuvette with two different compartments (sample and analysis) separated by a dialysis membrane (12 kDa molecular weight cut-off) that only allows the LEVO diffusion was employed for the experiments. Briefly, 170 μL of a suspension of the loaded MSN material in PBS (8 mg/mL) was placed in the sample compartment and 3.6 mL of fresh PBS were located in the analysis one. This system was kept at 37 °C and magnetic stirring was applied to the analysis compartment. At every analysis time, the solution in the analysis compartment was removed for analysis and replaced by fresh medium. The amount of LEVO released was determined by fluorescence spectroscopy using a BiotekPowerwave XS spectrofluorimeter, version 1.00.14 of the Gen5 program, with $\lambda_{ex}$ = 292 nm and $\lambda_{em}$ = 494 nm. First, a calibration curve was established in a concentration range from 0.01 to 12 μg/mL.

### 2.4. Microbiological assays

Briefly, bacteria culture was carried out by inoculation in Todd Hewitt broth (THB; Sigma-Aldrich) and incubated overnight at 37 °C with orbital shaking at 200 rpm. After culture, bacteria were centrifuged for 10 min at 3,500 g at 22 °C. The supernatant was then discarded and the pellet was washed three times with sterile PBS. The bacteria were then suspended and diluted in PBS to obtain a concentration of $2 \times 10^8$ bacteria per mL. Bacteria concentration was determined by spectrophotometry using a visible spectrophotometer (Photoanalizer D-105, Dinko instruments). Gram-negative bacteria as *Escherichia coli* (*E. coli* ATCC25922 laboratory strain) were used.

**2.4.1. Antimicrobial effect of levofloxacin dosage from MSNs**

To determine the effectiveness of the LEVO dosages released from the different MSNs against bacteria growth, 100 μL of each dosage was inoculated in 900 μL of $10^8$ bacteria per mL in PBS and incubated overnight. The presence or not of bacteria, as well as their quantification, was determined by counting the colony forming units (CFUs) in agar [35]. In this sense, 10 μL of this solution was seeded onto tryptic soy agar (TSA) and incubated at 37 °C overnight and subsequent counting. Controls containing bacteria was also performed and the experiments were performed in triplicate.

**2.4.2. Internalization into *E. coli* bacteria**

Sterilized round cover glasses were coated with poly-D-lysine (0.3 mL per well from a 0.1 mg/mL stock solution in Dulbecco's Phosphate Buffered Saline, PBS, Sigma Aldrich) for 90 min. Then, the excess poly-D-lysine was removed by washing with sterilized Milli-Q water (2 × 0.5 mL) and the round cover glasses were left to dry overnight in a sterile environment. The poly-D-lysine coated round cover glasses were placed in 24 well culture plates (CULTEK). Then, 500 μL of the $2 \times 10^8$ bacteria per mL was added onto the cover glass and subsequently 500 μL of the nanosystems suspensions in PBS were inoculated and incubated at 37 °C and 200 rpm in orbital stirring during 90 min. Different concentration of nanosystems (5, 10 and 20 μg/mL) were studied. Then, the glass discs were washed twice with sterile Hank's balanced salt solution (HBSS, Sigma-Aldrich) and 1 mL of HBSS was added. To stain the bacteria wall in red, FM 4-64FX dye (0.2 mL from a 5 μg/mL stock solution in HBSS, Invitrogen) was added and incubated on ice for 10 min with regular mixing. After incubation, the cover glasses were washed with HBSS and fixed with 2% wt/vol paraformaldehyde in PBS for 10 min at RT. The fixative was removed with HBSS twice and 0.5 mL of HBSS was added to each well before imaging. The samples were examined in an Olympus FV1200 confocal microscope. All images are representative of three independent experiments.

**2.4.3. Antimicrobial effect of MSNs and LEVO-loaded MSNs in planktonic bacteria**

To determine the antimicrobial effect of the different MSNs and LEVO-loaded-MSNs 10 µg/mL of each sample was inoculated in 900 µL of $10^8$ bacteria per mL in PBS and incubated during 90, 180 and 1440 min. For comparison, the antimicrobial effect of free LEVO (1.88 µg/mL) was also evaluated. The presence or not of bacteria, as well as their quantification, was determined by CFUs in agar. Thus, 10 µL of this solution was seeded onto TSA, incubated at 37 °C overnight and subsequently the number of CFUs was counted. Controls containing bacteria was also performed and the experiments were performed in triplicate.

**2.4.4. Antimicrobial effects against Gram-negative *E. coli* biofilm**

Effectiveness of the LEVO loaded nanosystems against biofilm was also determined. For this purpose, *E. coli* biofilms were previously developed onto poly-D-lysine coated round cover glasses by suspended the round cover glasses in a bacteria suspension of $10^8$ bacteria per mL during 48 h at 37 °C and orbital stirring at 100 rpm. In this case, the medium used was 66% THB + 0.2% glucose to promote a robust biofilm formation. After that, the round cover glasses containing the biofilm were placed onto 24 well culture plates (CULTEK) in 1.5 mL of new medium. Then, 0.5 mL of a suspension of MSN materials in PBS at final concentration of 5 and 10 μg/mL was added. After 90 min of incubation, the glass discs were washed three times with sterile PBS and 3 μL/mL of Live/Dead® Bacterial Viability Kit (Backlight$^{TM}$) was added. Also, 5 μL/mL of calcofluor solution was added to stain the mucopolysaccharides of the biofilm (extracellular matrix) in blue to specifically determine the biofilm formation. Both reactants were incubated for 15 min at RT. Controls containing biofilm bacteria were also performed. Biofilm formation was examined in an Olympus FV1200 confocal microscope and eight photographs (60× magnification) were taken of each sample [36,37]. The surface area covered with adhered bacteria was calculated using ImageJ software (National Institute of Health, Bethesda, MD). All images are representative of three independent experiments.

## 2.5. Statistics

Data are expressed as means ± standard deviations of a representative of three independent experiments carried out in triplicate. Statistical analysis was performed using the Statistical Package for the Social Sciences (SPSS) version 19 software. Statistical comparisons were made by analysis of variance (ANOVA). Scheffé test was used for post hoc evaluations of differences among groups. In all of the statistical evaluations, $p<0.05$ was considered as statistically significant.

## 3. Results and Discussion

### 3.1. MSN material synthesis

Fluorescent MSNs were synthetized by covalently linking the fluorescent dye to the silica network following a modified Stöber method [38-40]. SEM analysis of the obtained MSNs revealed a uniform spherical morphology with an average particle diameter of *ca.* 150 nm (Fig.S1 in Supporting Information). Low-angle XRD measurement showed a well-resolved pattern of 2D hexagonal MCM-41 materials, with a unit cell parameter of 4.55 nm. The well-ordered mesoporous structure was also confirmed by $N_2$ adsorption porosimetry, possessing an average pore diameter of 2.4 nm and 1090 $m^2/g$ of surface area.

### 3.2. External functionalization of the MSN material

The functionalization of MSN materials was intended only in the outer surface of the nanoparticles to keep free the inner mesoporous surface (Fig.1). The post-synthetic method *via* condensation of alkoxysilane derivatives with silanol groups of the silica surface, under water-free conditions, was used to covalently attach the amine moieties to the silica support [41]. The amount of the alkoxysilane was established from the surface area ($S_{BET}$) of the MSNs, of which the 25% was taken into account for the external surface, with an average surface concentration of Si-OH groups of 4.9 $OH/nm^2$ [42]. A molar ratio of three Si-OH groups with one R-Si(OEt)$_3$ moiety was considered as the stoichiometry of the condensation reaction.

To graft DAMO moieties to the outer surface of the MSNs (**MSN-DAMO**), the as-synthesized material containing the organic template filling the pores was used in the condensation reaction and its removal was performed afterwards [43]. By using this approach, a preferential functionalization of the external surface of MSNs is expected [43]. A 100% nominal degree of the external surface functionalization was considered to attain a maximum coverage.

However, for the preparation of the **MSN-G3** material, we kept in mind the possibility that the size of third generation dendrimers would prevent the functionalization of the interior of the MCM-41 pores [40]. In addition, the expected steric hindrance between third generation dendrimers at the silica surface prevents from a total silanol functionalization [33]. Thus, for the preparation of the **MSN-G3** material, we calculated the amount of dendrimer required to functionalize a quarter of the external surface of the MSNs.

### 3.3. Materials characterization

The organic content of the MSN materials was quantified from TGA and elemental chemical analyses to follow the incorporation of the alkoxysilane derivatives (Table 1). The results confirm the expected increase of the organic content after functionalization of the MSNs, although the **MSN-DAMO** material possesses a higher organic content compared to the dendritic functionalized **MSN-G3**. This fact reflects the steric hindrance that occurs when macromolecules are employed to functionalize the silica surface, resulting in a more effective incorporation of the smaller alkoxysilane. Therefore, in terms of functionalization degree (calculated in mmol per gram of material) the **MNS-DAMO** material incorporates almost 18 times more alkoxysilane reactive groups than **MSN-G3.**

Changes in the $\zeta$-potential values of the different nanoparticles were also used to assess the functionalization process. Incorporation of DAMO, as well as incorporation of the G3 dendrimer, produces a drastic change on the MSN material from negative to positive $\zeta$-potential values. As shown in Table 1, values change from −36.4 mV in bare MSN to *ca.* +35 mV in the hybrid materials. Both functionalized materials possess amine groups susceptible to protonation; however, the incorporation of the polycationic dendrimer does not exceed the positive value reached with the DAMO molecule. The possible acid-base equilibriums of the different functional groups over the nanoparticle silica surface in water and their corresponding p$K_a$ values, considering the acid-base properties of the different functional groups, are the following: deprotonation of the silanol groups in all MSN materials, the bare and the amino functionalized hybrid materials [Eq. (1)], protonation of the primary and secondary amine groups in DAMO [Eq. (2) and (3)] and protonation of the primary and tertiary amine groups in G3 dendrimer [Eq. (4) and (5)] [44-46].

$$R\text{-Si-OH} + H_2O \leftrightarrow R\text{-SiO}^- + H_3O^+ \qquad pK_a \approx 6.8 \qquad \text{Eq. (1)}$$

$$R\text{-NH}_2 + H_2O \leftrightarrow R\text{-NH}_3^+ + OH^- \qquad pK_a \approx 10.4 \qquad \text{Eq. (2)}$$

$$R_2\text{-NH} + H_2O \leftrightarrow R_2\text{-NH}_2^+ + OH^- \qquad pK_a \approx 9.2 \qquad \text{Eq. (3)}$$

$$R\text{-NH}_2 + H_2O \leftrightarrow R\text{-NH}_3^+ + OH^- \qquad pK_a \approx 9.8 \qquad \text{Eq. (4)}$$

$$R_3\text{-N} + H_2O \leftrightarrow R_3\text{-NH}^+ + OH^- \qquad pK_a \approx 6.1 \qquad \text{Eq. (5)}$$

Therefore, two main factors can be considered to explain a more positive value for the **MSN-DAMO** material: First, it is reflected the greater degree of functionalization achieved with the smaller molecule (see F value in Table 1). In **MSN-DAMO** most of the silanol groups (-Si-OH) over the nanoparticle surface have reacted during the functionalization, therefore not contributing to the -Si-O$^-$ formation in water (see also $^{29}$Si MAS NMR discussion below). On the other hand, the p$K_a$ values of amine moieties in G3 are slightly lower than p$K_a$ values of amines in DAMO, mainly due to a polyelectrolyte behavior in the macromolecule that reflects the effect of electrostatic repulsion between different charged sites. This effect entails the decrease of the p$K_a$ value with the increase of the protonation degree. Therefore, under the same conditions in the water media protonation would be more favorable for **MSN-DAMO** than for **MSN-G3** material. The same behavior was found in PBS medium; however, ζ-potential values measured in THB were found to be negative for bare **MSN** as well as both functionalized **MSN-DAMO** and **MSN-G3** materials, probably reflecting the adsorption of a protein corona on the surface of the nanoparticles [47].

The hydrodynamic particle diameter ($D_H$) of the nanoparticles was determined by DLS in water, PBS and THB media and the obtained values are shown in Table 1. The materials suspended in water display a hydrodynamic size distribution comprised between 140-190 nm, which indicates that the grafting process does not significantly alter the nanoparticle size. Nonetheless, the functionalized

MSNs possess a smaller hydrodynamic size, reflecting a decrease in the magnitude of the aggregates of nanoparticles in water media. All MSNs exhibit high enough ζ-potential values, either negative or positive, to be in the colloidal stability zone [48], but the presence of organic molecules attached to the external surface of the MSNs in the hybrid materials contributes as well to a steric repulsion that decreases the size of the aggregates in solution. In PBS and THB media the hydrodynamic size for all the materials increases slightly, being the maximum registered values 234 nm for MSN sample in THB and 263 nm for MSN-G3 sample in PBS. This slight increase in the hydrodynamic size for all materials in PBS and THB with respect to water may be a reflection of the shift in the ζ-potential toward values somewhat further away from the colloidal stability zone. However, $D_H$ values for all materials in the different media are in a range where large aggregates can be discarded.

Fig. 2A shows the $^1$H high-resolution magic angle spinning (HRMAS) NMR spectrum of the **MSN-G3** material. The chemical shifts closely match those of the respective functional groups in the solution NMR spectrum of the free G3 dendrimer, supporting the presence of this compound on the surface of the nanoparticles [33].

The presence of the G3 dendrimer on the surface of the nanoparticles is supported by the chemical shifts of the hybrid material, which closely match those of the respective functional groups in the solution NMR spectrum of the free compound [33]. The broadening of the signals with respect to the solution NMR spectrum is a normal fact due to the immovilization of the organic matter onto the solid nanoparticles. A sligth downfield shift of the signals can be adscried to the use of different deuterated solvents as well as techniques and equipments [44,49].

Fig. 2B shows the $^{13}$C{$^1$H} CP MAS NMR spectrum of the **MSNs-DAMO** material including the chemical shifts and their assignments. The observed $^{13}$C chemical shifts correspond to the DAMO molecule proving its presence in the nanoparticles.

A common fact for **MSN-G3** and **MNS-DAMO** materials is that signals due to ethoxy groups show up in both $^1$H and $^{13}$C MAS NMR spectra, respectively. These signals are also present in the $^{13}$C MAS NMR of the bare **MSN** sample. Furthermore, if we take into account that a 7.8% of organic content was found by TG analysis for the bare MSNs, the explanation for this signals is that they come from ethoxy groups from incomplete hydrolysis and condensation of the TEOS precursor during the sol-gel synthesis of MSNs. Then, the extraction method used to remove the CTAB surfactant leaves intact these groups in the silica network, unlike a calcination process would do.

Further analysis of the functionalization of the MSNs was performed by $^{29}$Si MAS NMR spectroscopy (Fig. 2C). The chemical shifts at around $\delta = -94, -102$ and $-112$ ppm apperar in all the spectra and represent the $Q^2$ [Si(OSi)$_2$(OX)$_2$], $Q^3$ [Si(OSi)$_3$(OX)] and $Q^4$ [Si(OSi)$_4$] silicon sites, respectively (X = H, C). Table 2 displays the populations of these silicon environments. The conversion of Si-OH groups to fully condensed Si-O-Si species during the post-grafting reaction is reflected in a decrease on the $Q^2$ and $Q^3$ peak areas and an increase in the $Q^4$ peak area, *i.e.*, a decrease in the $(Q^2 + Q^3)/Q^4$ ratio. This fact confirms the existence of covalent linkages between the silica surface and the organic groups in the **MSN-DAMO** and **MSN-G3** materials. Signals for $T^2$ [R-Si(OSi)$_2$(OX)] and $T^3$ [R-Si(OSi)$_3$] silicon sites appear at around $\delta = -58$ and $-68$ ppm, respectively, in the **MSN-DAMO** spectra. However, signals for the T groups in the spectra of the **MSN-G3** material are absent due to the low proportion of this silicon environment in the silylated G3 dendrimer.

The remaining populations of $Q^2$ and $Q^3$ sites after functionalization, for both **MSN-DAMO** and **MSN-G3** hybrid materials, indicates that the inner surface of the channels was preserved from functionalization. In the case of **MSN-DAMO**, this circumstance is due because the alkoxysilane grafting was done in advance to the surfactant extraction. In the case of **MSN-G3**, the fact is adscribed to the bulky size of G3 dendrimers that leads to the attachment of the macromolecules onto the external surface of the MSNs. These findings are also verified by the N$_2$ sorption studies detailed below (*vide infra*). The lower decrease in the $(Q^2+Q^3)/Q^4$ ratio for the **MSN-G3** sample also points to a lower coverage of the silica surface due to steric hindrance between third generation dendrimers.

Concerning the structural characterization of MSNs before and after functionalization, XRD confirms the 2D hexagonal mesoporous arrangement, showing three well-defined reflections that can be indexed as 10, 11 and 20 of a *p6mm* plain group for all the materials (Fig. 3A). These results agree with TEM studies (Fig. 3B-D), which show spherical shaped nanoparticles with an average diameter of *ca.* 120 nm. Regarding the mesoporous structure, TEM images evidence the presence of a honeycomb-like mesoporous arrangement typical of MCM-41 structure. These results reveal that neither the morphology nor the mesoporous order are affected by the used functionalization procedures.

Fig. 3E shows the $N_2$ adsorption isotherms corresponding to pristine **MSN**, **MSN-DAMO** and **MSN-G3** materials, respectively. All isotherms are IV type according to the IUPAC classification, which are characteristic of mesoporous materials exhibiting parallel cylindrical pores [50]. The appropriated treatment of the $N_2$ adsorption-desorption data allows to determine the main textural parameters (Table 3). Concerning the $S_{BET}$, it experiences a decrease in both functionalized samples compared to pristine MSN, because of the post-grafting of organic moieties [51]. However, the pore volume ($V_P$) remains unaltered in **MSN-G3** because, in good agreement with NMR results, due to the high volume of G3 this macromolecule is not able to functionalize the inner mesoporous cavities [33]. However, $V_P$ undergoes a remarkable drop in **MSN-DAMO**. This fact could be ascribed to the smaller size of DAMO, which may partially block the pore entrances. These findings are in good agreement with the calculated pore diameter ($D_P$) and wall thickness ($t_{wall}$) values, which remain constant in **MSN-G3** and experience a significant variation ($D_P$ decrease and $t_{wall}$ increase) in **MSN-DAMO**, compared to unmodified **MSN** material.

### 3.4. "In vial" levofloxacin release performance

LEVO was loaded into the different materials by forming well-dispersed suspensions of nanoparticles in a solution of LEVO in ethanol or $CH_2Cl_2$, *i.e.*, following an impregnation method (for further details see Experimental Section) [34]. The amount of LEVO loaded into the samples was 3.2%, 5.0% and 7.8% (wt.) for **MSN-L**, **MSN-DAMO-L**, **MSN-G3-L**, respectively. The effective drug loading was also evaluated by the reduction in the $V_P$ and pore diameter measured by $N_2$ adsorption (see Table S1 in Supporting Information). As above explained, the G3 macromolecule is not able to functionalize the inner mesoporous cavities and therefore the $V_P$ remains almost unaltered in **MSN-G3** compared to pristine **MSN**, being available more pore volume accessible for drug loading than in **MSN-DAMO**. In addition, the impregnation method followed for LEVO loading in **MSN-G3** is performed before the anchorage of the alkoxysilane on the external surface so the accessibility of LEVO to the inner pores is not impeded.

To assess the performance of the diverse nanosystems, namely **MSN-L**, **MSN-DAMO-L** and **MSN-G3-L**, under physiological relevant conditions (PBS 1x, pH = 7.4), LEVO release experiments were performed "in vial". Fig. 4A shows the LEVO release profiles from the different matrices after 14 days. Release profiles can be adjusted to first-order kinetic model by introducing an empirical non-ideality factor ($\delta$) to give the equation [52]:

$$Y = A(1 - e^{-kt})^\delta \qquad \text{Eq. (6)}$$

being Y the percentage of LEVO released at time *t*, *A* the maximum amount of drug released, and *k*, the release rate constant. In this approach, the values for $\delta$ are comprised between 1 for materials that obeys fist-order kinetics, and 0, for materials that release the loaded drug in the very initial time of test.

The parameters of the kinetics fitting are also shown in Fig. 4A indicate that the fastest LEVO release occurs from **MSN-G3** matrix ($k$ = 0.239 h$^{-1}$), exhibiting a $\delta$ = 1, which indicates a first-order release kinetics, and releasing 100% of loaded LEVO after 72 h of assay. In the case of **MSN-DAMO** and **MSN** there is a slight variation from the ideal first-order kinetics, with comparable $\delta$ values in the 0.6-0.7 range. In this case, release rate of **MSN-DAMO** is faster than that of **MSN**, with $k$ values of 0.051 h$^{-1}$ and 0.021 h$^{-1}$, respectively. Previous studies have revealed that there is a strong interaction between the LEVO molecules and silica network in MSNs *via* hydrogen bonding between *zwitterionic* form of this quinolone antibiotic at pH 7.4 and –Si-OH groups of silica nanoparticles [34,53]. Thus, pristine **MSN** sample partially retains the loaded LEVO, releasing less than 50% after 14 days of test. On the contrary, the presence of protonated amino groups on the external surface of both **MSN-DAMO** and **MSN-G3** samples promotes the interaction with LEVO molecules, which provokes drug departure from the mesopores resulting in a faster antibiotic release. Fig. 4B displays the LEVO dosages from different samples in the initial stages of the "in vial" assays, evidencing the significant highest antibiotic release from **MSN-G3**, which is 10-fold the minimum inhibitory concentration (MIC) for *E. coli* [54]. Likewise, the LEVO dosages released from **MSN-DAMO** are also significantly above the MIC, being the pristine **MSN** sample that exhibiting the lowest drug dosages albeit also above the MIC. The *in vitro* antimicrobial tests of these dosages *vs* Gram-negative *E. coli* reveal that they are effective for **MSN-G3** and **MSN-DAMO** after 1 and 3 h, observing bacterial growth for pristine **MSN**.

### 3.5. Evaluation of bacterial internalization

To directly observing the internalization capability on bacteria cells confocal microscope studies were conducted. Fig. 5 displays the confocal images of *E. coli* (labelled red with lipid membrane FM4-64FX) incubated during 90 min with pristine **MSN**, **MSN-DAMO** and **MSN-G3** (whose silica networks were labeled in green with fluorescein). Zoom of the confocal images at 60x of magnification clearly evidence notables different among all samples. All images show the bacillus morphology of *E. coli* with the bacterial wall labeled in red, in good agreement with the control. In the case of pristine **MSN**, green scattered dots not close to the bacteria walls are observed, which can be attributed to aggregates of nanoparticles not removed during the washing procedure and do not interacting with the bacteria wall. On the contrary, **MSN-DAMO** images display small domains of green aggregates corresponding to nanoparticles in the proximity of the bacterial wall obeying certain interactions between positive charges of amine groups and the negative bacterial wall. Nonetheless, a complete internalization and uniform localization of **MSN-G3** nanoparticles throughout the cytosol of the bacteria cells is appreciated. These results have been also observed for lower concentrations of MSNs of 5 µg/mL (see Fig. S2A in Supporting Information). As bacteria are negatively charged and both, **MSN-DAMO** and **MSN-G3** nanoparticles have positive charge density, electrostatic interactions bring them into contact with each other. Difference in the successful internalization of both hybrid materials may be attributed to the G3 dendritic skeleton, which provides **MSN-G3** of high surface flexibility. Consequently, there are more interaction points in the positively charged amine functional groups of **MSN-G3** that are exposed to the negatively charged bacterial wall [10]. Therefore, **MSN-G3** not only binds to the negatively charged phospholipid membranes, but also provokes a slight change in membrane permeability, which finally leads to nanosystem internalization into the bacteria, as it has been reported elsewhere for other systems [13,55]. Further studies to analyze the effect of **MSN-G3** internalization in the bacteria at longer times, as well as to elucidate the interaction and internalization between the nanosystem here proposed and bacteria cell membrane are being conducted for a better understanding of the mechanism against Gram-negative bacteria. Moreover, to evaluate the potential antimicrobial effect of the different samples (**MSN**, **MSN-DAMO** and **MSN-G3**), preliminary studies were carried out by incubation of 10μg/mL of each material in $10^8$ bacteria/mL suspensions during 90, 180 and 1440 min, followed by plating and counting of CFUs (Fig. S3A in Supporting Information). The obtained results reveal no significant antimicrobial effect for all materials after 90 min of assay, in good agreement with confocal microscopy images (Fig. 5). However, after 180 min of assay there is a significant antimicrobial effect in **MSN-DAMO** and **MSN-G3** compared to the control. Finally, after 24 h of incubation, the highest antimicrobial effect corresponds to **MSN-G3**, being significantly higher than the control and the other tested samples, reducing *ca.* 99% the number of CFUs. In addition, to further investigating the effect of the incorporation of LEVO into the different MSNs in comparison to free LEVO, complementary antimicrobial assays were also performed. The results displayed in Fig. S3B (in Supporting Information) indicate that in all cases the presence of LEVO leads to a noticeable decrease in the number of CFUs and that this effect is more remarkable with the increase of the incubation time.

### 3.6. *In vitro* antimicrobial efficacy against *E. coli* biofilms

After evaluating the internalization effectiveness and antibiotic release profiles, *in vitro* antimicrobial assays can be performed to assess the efficacy of these nanosystems against *E. coli* biofilms. Fig. 6 shows the most representative results derived from the effects of **MSN-L**, **MSN-DAMO-L** and **MSN-G3-L** suspensions onto the preformed *E. coli* biofilm by confocal microscopy. Initially, the preformed biofilm displays a typical structure composed of colony live bacteria (green) covered by a protective mucopolysaccharide matrix (blue). After 90 min of incubation with the different nanoparticles, remarkable differences are detected. **MSN-L** sample is not able to destroy the biofilm, appearing live bacteria colonies coated with its protective layer. However, small amounts of scattered dead bacteria are present onto the outermost surface probably due to the action of the LEVO released from this material, in agreement with Fig. 4. On the other hand, **MSN-DAMO** triggers the biofilm disruption, with the presence of small unfilled domains within the colony, showing vastly live cells (green), dispersed mucopolysaccharide (blue) fragments and scarcely dead cells (red). Finally, the best results are derived from **MSN-G3-L** assays, which is able to totally destroy the biofilm since no blue staining and almost negligible live cells are observed. It should be mentioned that even at nanoparticles concentrations of 5 µg/mL the biofilm is destroyed, existing dead bacteria (red) covering almost the entire biofilm surface (Fig. S2B in Supporting Information). This finding agrees with the synergistic combination of the bacterial penetration capability of polycationic dendrimers present in **MSN-G3** (Fig. 5) and the antimicrobial effect of LEVO released from this matrix (Fig. 4).

In addition, control experiments with G3 dendrimer, LEVO and a combination G3 and LEVO were carried out (Fig. S4 in Supporting Information). The concentration of each compound was that corresponding to the G3 grafted to MSNs (MSN-G3) and to the amount of LEVO released from MSN-G3-L after 90 min during the "in vial" assay. The results evidence not significant antimicrobial effect in any of the cases with respect to the control.

## 4. Conclusions

A new class of antimicrobial agent, named "nanoantibiotic", based on mesoporous silica nanoparticles (MSNs) decorated with polypropyleneimine dendrimers of third generation (G3) and loaded with levofloxacin (LEVO) as antibiotic has been developed. The covalently grafting of these G3 dendrimers to MSNs (MSN-G3) allows an effective internalization in Gram-negative bacteria. This internalization can be ascribed to the high density of positive charges and flexibility on the surface of the MSN-G3, which produces electrostatic interactions with the negatively-charged bacteria walls and triggers their permeabilization. Furthermore, the LEVO antibiotic loaded into the mesoporous cavities is released in a sustained manner at effective antimicrobial dosages. These findings demonstrate that the synergistic combination of bacterial internalization and antimicrobial agents into a unique nanosystem provokes a remarkable antimicrobial effect against bacterial biofilm. This novel nanoantibiotic is envisioned as a promising alternative to the current available treatments for the management of infection.

## Appendix A. Supplementary data

Supplementary data associated with this article can be found in the online version.

## Disclosure section

The authors declare not conflict of interest.

## Acknowledgements

This work was supported by European Research Council, ERC-2015-AdG (VERDI), Proposal No. 694160; Ministerio de Economía y Competitividad (MINECO) grants MAT2013-43299-R, MAT2015-64831-R and MAT2016-75611-R AEI/FEDER; Instituto de Salud Carlos III (ISCIII) grant PI15/00978, and FEDER funds of the European Union. CIBER is a public research consortium created by ISCIII whose actions are co-funded by the European Regional Development Fund.

# Figure Captions

**Fig. 1** Synthetic routes used for the external functionalization of MSNs affording A) **MSN-DAMO** and B) **MSN-G3** materials.

**Fig. 2** A) $^1$H HRMAS NMR spectrum (suspension in D$_2$O) of the **MSN-G3** material. B) $^{13}$C{$^1$H} CP MAS NMR spectra of the **MSN** and **MSN-DAMO** materials. The assignation of the signals in the MSN-DAMO spectrum is depicted in the representative structure of the material. The assignments are somewhat ambiguous for C$_4$ and C$_5$ carbons, *i.e.* they may be reversed. C) $^{29}$Si MAS NMR spectra of the MSN materials synthetized in this work. For A) and B) signals due to ethoxy groups from the incomplete hydrolysis and condensation during the sol-gel synthesis of the MSNs are indicated with a # mark. Residual methoxy groups from DAMO are also contemplated. As well, the signals due to residual surfactant in the materials are indicated with a * mark in the $^{13}$C spectrum.

**Fig. 3** A) Low angle powder X-ray diffraction patterns of the mesoporous materials synthesized in this work. Reflections of a 2D hexagonal (*p6mm*) plane array of pores are indexed in the **MSN** XRD pattern. TEM images corresponding to B) **MSN**, C) **MSN-DAMO** and D) **MSN-G3** materials. E) N$_2$ adsorption-desorption isotherms of the MSN materials, before and after functionalization with DAMO and the G3 dendrimer.

**Fig. 4.** A) "In vial" cumulative LEVO release profiles from **MSN-L**, **MSN-DAMO-L** and **MSN-G3-L** nanosystems. B) LEVO dosages released from the three loaded MSN materials after 1 h and 3 h. *In vitro* antimicrobial assays of each dosage have been carried out in *E. coli* cultures. ⊕ symbol indicates the bacterial growth of CFUs in the culture, indicating the infectivity of dosage. On the contrary, ⊖ symbol represents the inhibition of the bacterial growth with no CFUs observed for these dosages.

**Fig. 5.** Confocal microscopy images of Gram-negative *E. coli* bacteria after treatment with pristine **MSN**, **MSN-DAMO** and **MSN-G3** materials at 10 μg/mL during an incubation time of 90 min. The *E. coli* control culture after 90 min without treatment is also displayed. The *E. coli* cell membrane was stained with FM4-64FX (red) and the MSN materials have previously been fluorescently tagged with fluorescein (green). Scale bars, 3 μm.

**Fig. 6.** Confocal microscopy study of the antimicrobial activity of the LEVO loaded MSNs materials onto Gram-negative *E. coli* biofilm. The confocal images show the biofilm preformed onto covered glass-disk after 90 min without treatment (control) and after 90 min of incubation with **MSN-L**, **MSN-DAMO-L** and **MSN-G3-L** at 10 μg/mL. Live bacteria are stained in green, dead bacteria in red and the protective biolayer, *i.e.*, the extracellular polysaccharide matrix biofilm in blue. Scale bars, 20 μm.

Figure 1

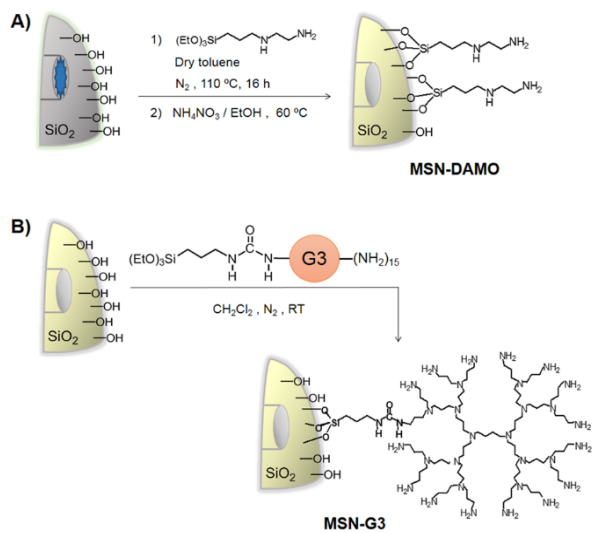

Figure 2

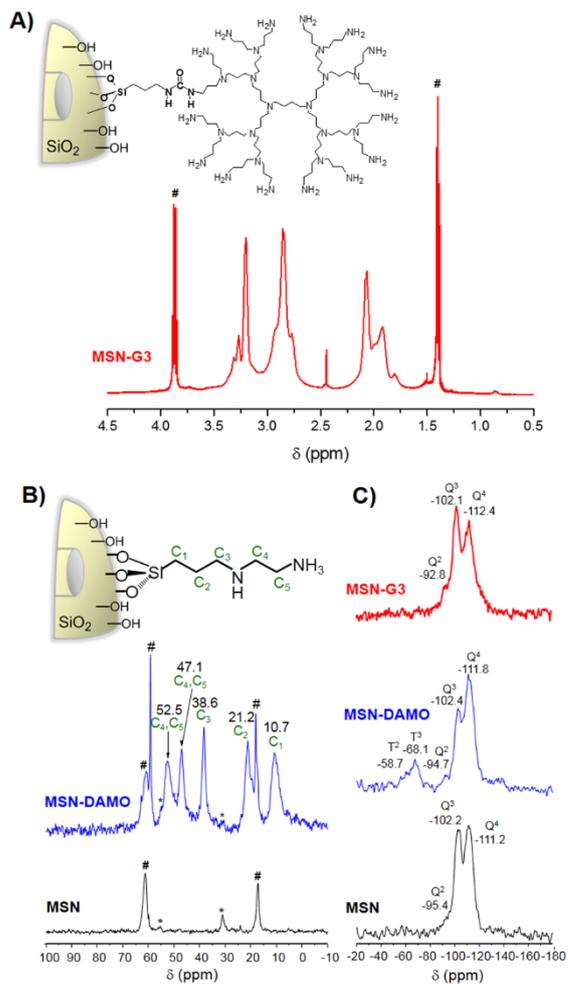

Figure 3

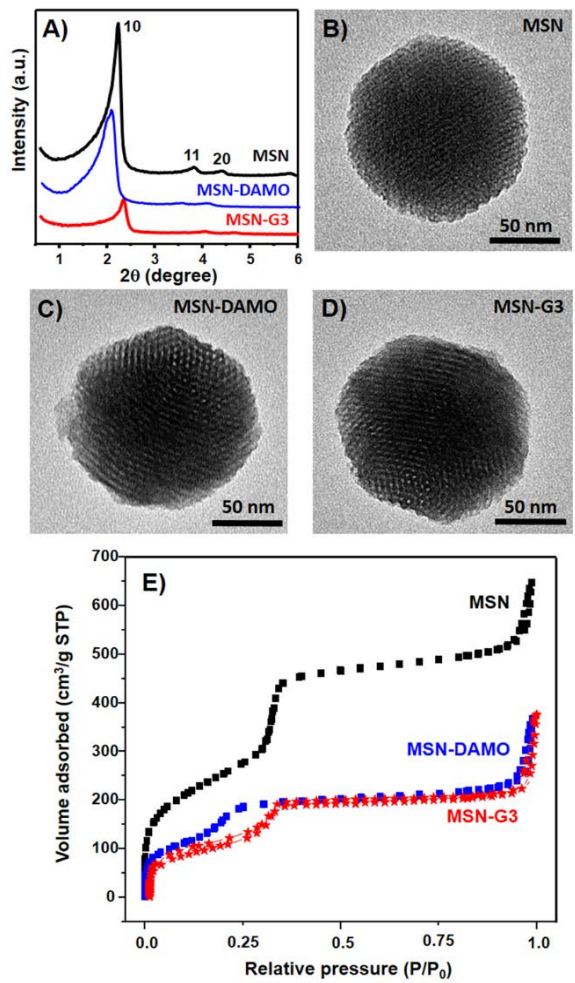

Figure 4

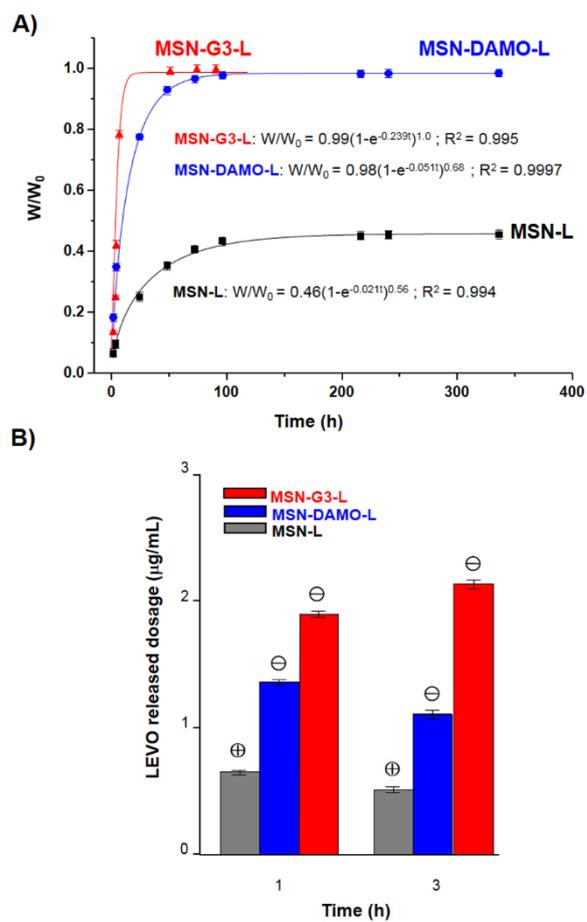

Figure 5

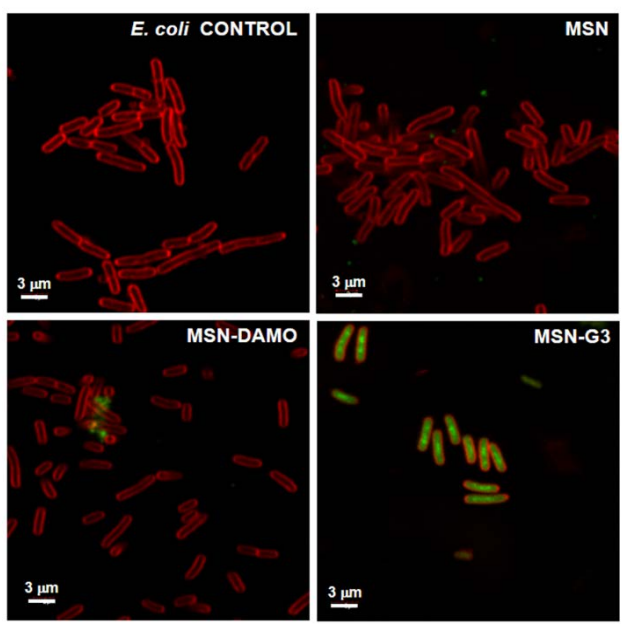

Figure 6

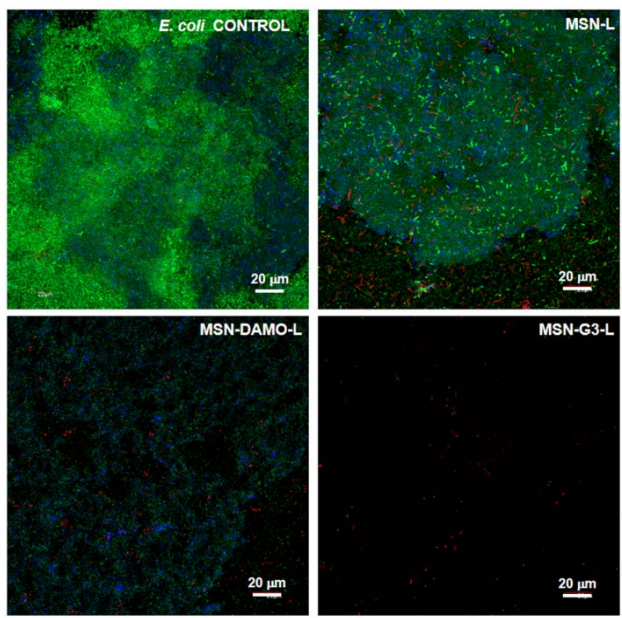

Graphical abstract

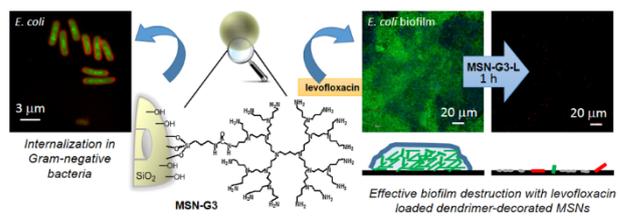